\documentstyle[aps]{revtex}
\input epsf
\epsfverbosetrue

\begin{document}
\draft
\sloppy
\twocolumn[
\hsize\textwidth\columnwidth\hsize\csname
@twocolumnfalse\endcsname

\title{Shot Noise in Schottky's Vacuum Tube is Classical}

\author{C.~Sch\"onenberger and S.~Oberholzer}
\address{Physics Departement, University of Basel,
Klingelbergstr.~82, CH-4056 Basel, Switzerland}

\author{E.~V.~Sukhorukov}
\address{Department of Physics, University of Florida,
Gainesville, FL 32611-8440, USA}

\author{H.~Grabert} \address{Faculty of Physics,
University of Freiburg, Hermann-Herder-Strasse 3, D-79104
Freiburg, Germany}

\date{\today}
\maketitle

\begin{abstract}
In these notes we discuss the origin of shot noise
('Schroteffekt') of vacuum tubes in detail. It will be shown that
shot noise observed in vacuum tubes and first described by W.
Schottky in 1918 \cite{Schottky1918} is a purely classical
phenomenon. This is in pronounced contrast to shot noise
investigated in mesoscopic conductors \cite{BlanterReview1999}
which occurs due to quantum mechanical diffraction of the
electronic wave function. \pacs{84.47+w, 72.70.+m}
\end{abstract}
]

\section{Introduction}
Shot noise is due to time-dependent fluctuations in the electrical
current caused by the random transfer of discrete units of
charge. In 1918 Schottky \cite{Schottky1918} analyzed these
fluctuations in vacuum tubes theoretically for the first time
arriving at his
famous Schottky formula. It states that the spectral density $S$ of
the fluctuations at `low' frequencies is proportional to the unit of
charge $e$ and to the mean electrical current $|I|$.

In recent
years shot noise of mesoscopic conductors has been
investigated extensively \cite{BlanterReview1999}. In these
systems shot noise is a $quantum$ phenomenon originating from
the diffraction of wave functions \cite{BeenakkerPRB1991}.
It is the quantum mechanical uncertainty of not knowing with
absolute certainty whether a particle incident on a
scattering region transmit from source to drain that is
responsible for shot noise. Schottky derived his formula
before the existence of quantum mechanics, solely making use of
classical statistical mechanics. In retrospect one may
ask the question whether shot noise in the vacuum tube
is classical or not. The answer is not straightforward as
many discussions with colleagues have shown. Most engineers,
for example, are convinced that shot noise is a classical
phenomenon altogether. The mesoscopic physics community, on the
other hand, tend to believe that shot noise in electrical
conductor is quantum in general. There are colleagues
who are in favor of a quantum-mechanical origin for shot noise
observed in the vacuum tube. This has motivated us to
analyze the randomness contributing to shot noise in vacuum
tubes in detail. It turns out that quantum diffraction
in the emission process can be neglected in vacuum tubes.
The main source of noise stems from the the $classical$ occupation
of electron states (i.e. the Boltzmann tail)
within the cathode. Hence, Schottky's vacuum tube is classical!

\section{Shot noise of a two-terminal conductor}
We start with a simple derivation of the expression for the power
spectral density of the noise of a two-terminal conductor along
the lines of Martin and Landauer \cite{MartinPRB1992}. In their paper the
fluctuating currents are a result of the random
transmission of electrons from one terminal to the other.
Different processes contribute to the noise for each energy $E$
and mode $n$ (only elastic scattering is assumed):
\begin{enumerate}
    \item  A current pulse occurs whenever an electron wave packet
    incident from the left terminal (source) is scattered into an empty state
    in the right terminal (drain).
    The rate $\tau_{rl}^{-1}$ of these events is proportional to the probability
    $f_{L}(E)$ that an energy state $E$ in the left reservoir is
    occupied, times
    the probability \mbox{$1-f_{R}(E)$} that a respective state in the right
    reservoir is unoccupied,
    times the transmission probability from left to right:
    $T_{n}^{rl}(E)\equiv T_{n}(E)$:
\begin{equation}
    \tau_{rl}^{-1} \sim f_{L}(E)[1-f_{R}(E)]\,T_{n}(E).
\end{equation}

    \item  Of course the reverse process that electrons scatter from
    an occupied state in the right reservoir to an unoccupied state
    in the left reservoir contributes to noise, too. The rate of
    these processes is given by:
\begin{equation}
    \tau_{lr}^{-1}\sim f_{R}(E)[1-f_{L}(E)]\,T_{n}(E),
\end{equation}
    where $T_{n}^{lr}(E)=T_{n}^{rl}(E)\equiv T_{n}(E)$ has been
    taken into account.
\end{enumerate}
\noindent Since we require an expression for the fluctuations,
i.e. the deviations from the mean current, the mean current
squared has to be subtracted. As the current is proportional to
to $T_{n}(E)\,[f_{L}(E)-f_{R}(E)]$, the contribution from electrons at
energy $E$ in one specific mode $n$ to the noise is proportional
to:
\begin{eqnarray}
    &&f_{L}(E)[1-f_{R}(E)]\,T_{n}(E)
    +f_{R}(E)[1-f_{L}(E)]\,T_{n}(E)\rule[-2mm]{0mm}{5mm}\nonumber\\
    &&\hspace{18mm} - \,[f_{L}(E)-f_{R}(E)]^2\,T_{n}^2(E).
    \label{abcdef}
\end{eqnarray}
The coefficient follows from the fact that if no bias is applied
($V=0$) the expression for thermal noise $4k_{B}\theta G$ must be
recovered ($\theta$ is the temperature):
\begin{eqnarray}
    S & = &     4k_{B}\theta G \; =\; 4k_{B}\theta \,G_{0}\int dE\,
            \left(-\frac{\partial f}{\partial
            E}\right)\sum_{n}T_{n}(E) \nonumber \\
      & = &     2G_{0}\sum_n \int dE\,2f(E)[1-f(E)]T_{n}(E),
    \label{abcd}
\end{eqnarray}
with $G_{0}=2e^2/h$. In equilibrium ($V=0$),
$f_{L}(E)=f_{R}(E)\equiv f(E)$, where $f$ denotes the
Fermi-Dirac distribution. Then, expression
(\ref{abcdef}) equals
\begin{equation}
    2f(E)[1-f(E)]\,T_{n}(E).
    \label{abc}
\end{equation}
Comparing Eq.~(\ref{abcd}) and Eq.~(\ref{abc}) the general
expression for the shot noise of a two-terminal conductor (in the
zero frequency limit) follows as
\cite{LesovikJETP1989,ButtikerPRL1990}:
\begin{eqnarray}
    S & = & 2\,G_{0}\sum_{n}\int dE\,
        \{ f_{L}(1-f_{R})\,T_{n}+f_{R}(1-f_{L})\,T_{n}\nonumber\\
        && -\, [f_{L}-f_{R}]^2\,T_{n}^2\}\nonumber\\
    & = & 2\,G_{0} \sum_{n}\int dE\,
    \{ [f_{L}(1-f_{R}) + f_{R}(1-f_{L})]\,T_{n}(1-T_{n})\nonumber\\
    && +\,[f_{L}(1-f_{L})+f_{R}(1-f_{R})]^2\,T_{n}^2\}
    \label{xxxx}
\end{eqnarray}

\section{Vacuum tubes}
\begin{figure}[htb]
\centering \epsfxsize=76 mm \epsfbox{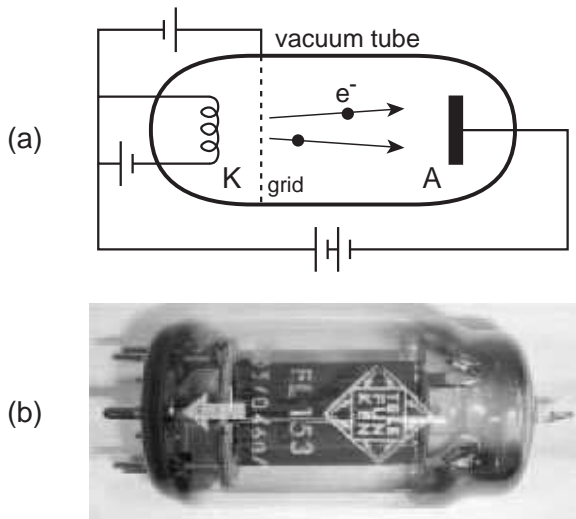} \vspace{8 pt}
\caption{Vacuum tubes: (a) Schematics of a triode.  Electrons
having enegies larger than the work function $W$ of the tungsten
filament are
emitted from the heated cathode (K), travel through the vacuum and
are attracted by the positive anode (A). (b) Photograph of a
historical tetrode (triode with additional grid) containing 4
electrodes (Telefunken EL 153).} \label{roehre}
\end{figure} \noindent
Figure~\ref{roehre}(a) shows a schematics of a vacuum tube
(triode): The heated cathode (K) made of a wounded tungsten wire
boils off electrons into the vacuum. The emitted elctrons
are attracted by the
positively charged anode (Edison effect). A negatively biased grid
(or many grids) between cathode and anode
controls the electron current. By designing the cathode, grid(s)
and plate properly, the tube converts a small AC voltage
at the grid
into a larger AC signal, thus amplifying it \cite{Koller1937}.
In the following we diregard the grid and consider only the
vacuum diode (as Schottky did).

\begin{figure}[htb]
\centering \epsfxsize=70 mm \epsfbox{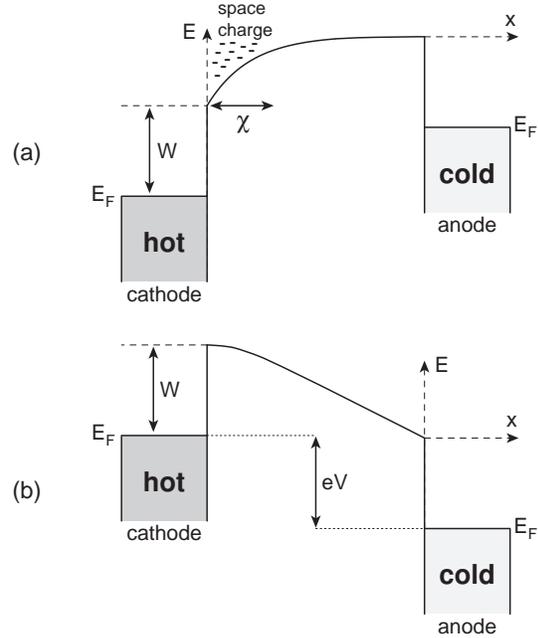} \vspace{8 pt}
\caption{(a) Space-charge region formed in front 
of the cathode in a open-circuited tube. (b) For sufficiently high
bias voltages $V$ the space-charge is removed (saturation regime)
and the potential drops linearly. $W$ denotes the work function.}
\label{diode}
\end{figure} \noindent
If the anode is floating, no net current flows from
cathode to anode [Fig.~\ref{diode}(a)]. Instead, a negative
space-charge is formed in front of the cathode, originating from
evaporated electrons which are hold back by the ionized atoms. The
size $\chi$ of the space-charge region can be calculated solving
the Poisson-equation $\Delta \varphi(x) = - en(x)/\epsilon_{0}$
for the electrical potential $\varphi(x)$ with the electron
density $n(x)=n_{0}\exp(-e\varphi(x)/k_{B}\theta) \simeq
n_{0}\,[1-e\varphi(x)/k_{B}\theta]$, where $n_{0}$ is the electron
densitiy within the cathode:
\begin{equation}
    \chi=\sqrt{\frac{\epsilon_{0}k_{B}\theta}{n_{0}e^2}}.
    \label{spacechargesize}
\end{equation}
The higher the temperature $\theta$
the larger the space-charge region.

\begin{figure}[h]
\centering \epsfxsize=60 mm \epsfbox{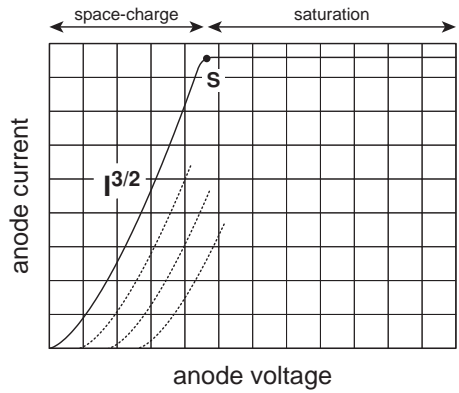} \vspace{8 pt}
\caption{Current-voltage characteristics of a vacuum tube
illustrating the 3/2-power law [Eq.~(\ref{powerlaw})] and the
saturation point (S). The dashed curves are IV-curves for
different grid voltages. Within the saturation regime the current
does no longer depend on the anode voltage because all electrons
emitted by the cathode are collected at the anode.}
\label{kennlinie}
\end{figure}

On the other hand, if the circuit is closed and the cathode is kept
at an elevated temperature, a thermionic current flows from
cathode  to anode [Fig.~\ref{diode}(b)]. The magnitude of this current is
limited by the negative space-charge region in front of the cathode.
This is also true if the anode is kept at a moderate positive potential.
In this space-charge limited regime,
the magnitude of the current is given by:
\begin{equation}
    I=\frac{\sqrt{2}}{9\pi}\sqrt{\frac{e}{m}}\frac{V^{3/2}}{L^2}
    \label{powerlaw}
\end{equation}
with $L$ the distance between cathode and anode \cite{Koller1937}.
Only for sufficiently large bias voltages $V$ are all
emitted electrons attracted by the anode
and the space-charge region is removed.
In this case, the current saturates (does no longer depend on the
anode voltage) and is determined by the temperature of the cathode
[Fig.~\ref{kennlinie}].
 \noindent

In the space-charge limited regime where the possibility of escape
of an electron is limited by Coulomb repulsion
shot noise is suppressed. Full shot
noise $S=S_{Poisson}=2e|I|$ is only present in the saturation
regime \cite{HullPhysRev1925}. The question whether shot noise
in the saturation regime is classical or quantum in nature will be
discussed in sect.~\ref{schrottefektintubes}. Before, the
electrical field and current in the saturation regime will be
estimated.

\section{Electrical field and current in the saturation regime}
At the edge of the vacuum barrier the electron density is
approximatively given by $n_{0}=a^{-3}\,\exp
(-W/k_{B}\theta)\simeq 5\cdot 10^{16}$ m$^3$ with $a\simeq 1.2$
$\AA$ the typical interatomic distance, $W= 4.5$ eV the work
function of tungsten and  $\theta=$ 2000 K the cathode temperature.
The size $\chi$ of the space-charge region follows from
(\ref{spacechargesize}) and is of the order 10 $\mu$m. The charge
build up at the cathode corresponds to an electrostatic surface
potential of $k_{B}\theta/e$, so that the surface electric field
can be estimated as $\mbox{$\cal{E}$}\simeq k_{B}\theta/e\chi$.
Inserting numbers the \emph{saturation field} is of the order
$10^4$ V/m.

The electrical current density due to thermionic emission from a
heated conductor is given by the Richardson-Dushman equation
\cite{SommerfeldBethe1967}:
\begin{equation}
    j=\mbox{$\cal{L}$}\,\theta^2\exp(-W/k_{B}\theta)
\end{equation}
with $\mbox{$\cal{L}$}=emk_{B}^2/2\pi^2\hbar^3=120$
AK$^{-2}$cm$^{-2}$. This expression is only correct if the
electrical field \mbox{$\cal{E}$} is so high that the
space-charge region is removed (saturation regime).

\begin{figure}[h]
\centering \epsfxsize=70 mm \epsfbox{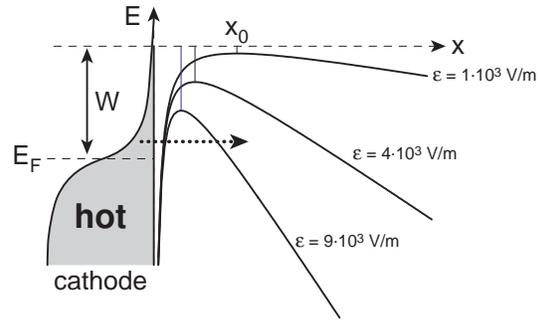} \vspace{8pt}
\caption{Image potential determining the barrier shape in emission
of electrons from the hot cathode for different electrical fields
$\mbox{$\cal{E}$}$. For very high fields the barrier becomes very
thin allowing for electrons to tunnel (quantum regime).} \label{diode3}
\end{figure} \noindent

An emitting electron experiences a potential $\phi$
determined by the saturation field
\mbox{$\cal{E}$} and its image potential
in the (planar) cathode (image-potential), Fig.~\ref{diode3}:
\begin{equation}
    \phi(x)=-\mbox{$\cal{E}$} x -\frac{e}{4\pi\epsilon_{0}}\frac{1}{x}
    \label{potential}
\end{equation}
The maximum of
$\phi(x)$ lies at $x_{0}=
\sqrt{e/4\pi\epsilon_{0}\mbox{$\cal{E}$}}$, where the barrier is
lowered by
$e\phi(x_{0})=-2e\sqrt{e\mbox{$\cal{E}$}/4\pi\epsilon_{0}}\simeq
-8$ meV. This is negligible in comparison with the work function
$W=4.5$ eV, so that the saturation current can be estimated
disregarding the barrier lowering. For a cathode area of 10$^{-2}$
cm$^{-2}$ and $\theta=2000$ K the \emph{emission current} is of
the order 10 $\mu$A.

\section{The `Schroteffekt' in vacuum tubes}
\label{schrottefektintubes} In the saturation regime (no
space-charge at the cathode) the shot noise
due to the emission of electrons from cathode to anode is according
to Eq.~(\ref{xxxx}) given by
\begin{equation}
    S= 2 G_{0}\sum_{n}\int dE\,
    \{ \underbrace{f_{cathode}T_{n}(1-T_{n})}_{quantum} +
    \underbrace{f_{cathode}\,T_{n}^2}_{classical} \}.
    \label{vakuumroehrenrauschen}
\end{equation}
Here we made use of the fact that
$f_R=f_{anode}=0$ and that
the occupation of states within the hot
cathode at energy $W$ above $E_F$ is small (classical):
$f_{L}=f_{cathode}=\exp
(-W/k_{B}\theta)\ll 1$. Therefore, $f_{L}(1-f_{L})\simeq f_{L}$.
The current $I$ due to emission at the cathode equals
\begin{equation}
    I = \frac{2e}{h}\sum_{n}\int dE\,f_{cathode}\,T_{n}
    \label{current}
\end{equation}

There are two terms in Eq.~(\ref{vakuumroehrenrauschen})
contributing to shot noise: the first term is of
quantum mechanical origin since it only contributes for transmission
probabilities $T\neq 0,1$, hence, only if there are quantum
uncertainties. In contrast, the second term is classically because
it dominates when all transmission
coefficients are classical, i.e. either $0$ or $1$.
As we show now, both classical and quantum parts may yield Schottky's famous result
independently.

\noindent
\textbf{Classical part:} 
Because all $T_{n}$ are either $0$ or $1$,
$T_{n}^2=T_{n}$ and shot noise is consequently
given by
\begin{equation}
    S  = 2 G_{0}\sum_{n}\int
    dE\,f_{cathode}\,T_{n}.
\end{equation}
With Eq.~\ref{current}
we arrive at Schottky's formula $S=2e|I|$.\\

\noindent \textbf{Quantum part:} In the special limit
in which all $T_{n}$'s are small ($T_{n}\ll
1$) the noise is due to tunneling (quantum diffraction).
In this
case the quantum term $\sim T_{n}$ in
Eq.~(\ref{vakuumroehrenrauschen}) dominates, while terms
proportional to $T_{n}^2$ are negligibly small.
Hence again, $S$ is given by
\begin{equation}
    S  = 2 G_{0}\sum_{n}\int dE\,f_{cathode}\,T_{n}.
\end{equation}
and we obtain Schottky's formula $S=2e|I|$, this time however,
originating from quantum diffraction.\\

\vspace{5mm}

In order to decide whether shot noise in vacuum
tubes is classical or quantum, the transmission probabilities
$T_{n}$'s need to be evaluated.

\section{Transmission probability at the cathode}

The quantum-mechanical transmission probability
for electrons with energy $\epsilon$ above the barrier
can be estimated with the following equation
\cite{Landau1965}:
\begin{equation}
    T\simeq\left[1+e^{-2\pi \epsilon/\hbar\omega_{0}}\right]^{-1}
    \label{transmissiontube}
\end{equation}
$\epsilon$ is of order $k_{B}\theta$ with $\theta$
the cathode temperature. $\omega_{0}$ denotes the
negative curvature at the barrier top and determines whether the
barrier is sharp or smooth. It can be obtained from the
`force-constant'
\begin{equation}
    f=e\left|\frac{\partial^2\varphi}{\partial x^2}\right|_{x=x_{0}}
    =2\sqrt{4\pi\epsilon_{0}e}\cdot\mbox{$\cal{E}$}^{3/2}
\end{equation}
with $\omega_{0}=\sqrt{f/m}$  \cite{comment1}:
\begin{equation}
    \omega_{0}=\left(\frac{16\pi\epsilon_{0}e}{m^2}\right)^{1/4}\cdot
    \mbox{$\cal{E}$}^{3/4}.
\end{equation}
If $\hbar\omega_{0}\ll k_{B}\theta$, $T=1$ and the classical part
of the shot noise in (\ref{vakuumroehrenrauschen}) dominates. In
the opposite limit $\hbar\omega_{0}\gg k_{B}\theta$, the transmission
probability $T$ is small and shot noise is due to (quantum-mechanical)
tunneling.

\begin{figure}[h]
\centering \epsfxsize=60 mm \epsfbox{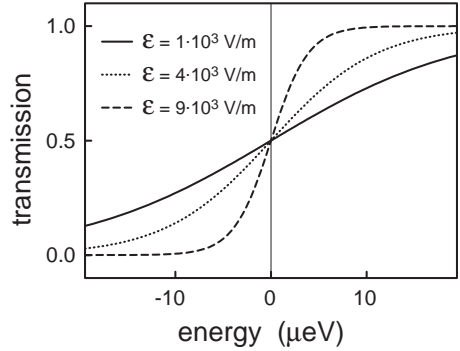} \vspace{8pt}
\caption{Transmission probability vs. energy $\epsilon$ according
Eq.~(\ref{transmissiontube}) for different saturation fields
\mbox{$\cal{E}$}.} \label{diode4}
\end{figure} \noindent

The first shot noise measurements in vacuum tubes were carried
out by Hartmann in 1921 \cite{HartmannAnnPhys1921}. A very careful
study of the `Schroteffekt' was performed by Hull and Williams in
1925 \cite{HullPhysRev1925}. In the first part of that latter
experiment shot noise was measured in
the saturation regime, where the thermionic current is limited by
temperature \cite{comment2}. The corresponding parameters are
given in the first two lines of Tab.~\ref{vacuumtubeparameters}.
In this regime the full Schottky-noise $2e|I|$ has been
measured in excellent agreement with Millikan's value for the
electron charge $e$. The ratio $\hbar\omega_{0}/k_{B}\theta\ll 1$,
so that the transmission probability is $1$. Therefore, shot noise observed
in this experiment is \emph{classical}.

\begin{table}[h]
\centering
\begin{tabular}{|c|c|c|c|c|c|c|c|}
\mbox{$\cal{E}$} [V/m] & $V_{G}$ [V] & $V_{P}$ [V] & $i_{0}$ [mA]
& $\theta$ [K] &
$\hbar\omega_{0}/k_{B}\theta$ & T & $F$ \rule[-2mm]{0mm}{6mm}\\
\hline
$3\cdot 10^6$ & 120 & 120 & 1 & 1675 & $3.2\,10^{-2}$ & 1 & 1.00 \rule[-1.5mm]{0mm}{5mm}\\
$3\cdot 10^6$ & 120 & 120 & 5 & 1940 & $2.7\,10^{-2}$ & 1 & 1.00 \rule[-1.5mm]{0mm}{5mm}\\
\hline
$1\cdot 10^4$ & -6  & 130 & 1 & 1675 & $4.4\,10^{-4}$ & 1 & 0.93 \rule[-1.5mm]{0mm}{5mm}\\
$1\cdot 10^4$ & -6  & 130 & 3 & 1805 & $4.1\,10^{-4}$ & 1 & 0.49 \rule[-1.5mm]{0mm}{5mm}\\
$1\cdot 10^4$ & -6  & 130 & 5 & 1940 & $3.8\,10^{-4}$ & 1 & 0.20 \rule[-2mm]{0mm}{5.5mm}\\
\end{tabular} \vspace{8pt}
\caption{Experimental parameters from shot noise measurements of
Hull and Williams in 1925. $V_{G}$ is the voltage at the grid and
$V_{P}$ at the anode plate. $i_{0}$ is the thermionic current.
$F=S/2e|I|$ denotes the Fano factor. The second last column shows
that the shot noise observed in this experiment is a classical
phenomenon.} \label{vacuumtubeparameters}
\end{table}\noindent

In the second part of the experiment the effect of the
space-charge on the shot noise was investigated at lower electric
fields \mbox{$\cal{E}$}. The corresponding parameters are given in
the last three lines of Tab.~\ref{vacuumtubeparameters}. At lower
temperatures the emission current is limited by temperature and
the full Schottky-noise is observed. At higher temperatures,
however, the space-charge builds up and shot noise is
gradually suppressed due to Coulomb interaction.
Note however, that for all these experiments $\hbar\omega_0/k_B\theta$
is a very small parameter, so that $T=1$ within very high accuracy.
Quantum corrections arise only in the $80$th decimal after
the comma!

\section{Conclusion and Acknowledgment}

In conclusion, we have shown (hopefully unambiguously) that shot noise in
vacuum tubes is in general \emph{classical}. This is in profound contrast
to shot noise observed in mesoscopic conductors.

This work was supported by the Swiss National Science Foundation
and the Institute for Theoretical Physics (ITP) at UCSB.



\begin{thebibliography}{10}
\bibitem{comment1}
This is analogous to the mechanical oscillator
$\ddot{x}+\omega_{0}^2 x=0$ with $F=-fx$ and
$E_{pot}=\frac{1}{2}fx^2$.
\bibitem{comment2}
In order to avoid space-charge a very strong electric field at the
cathode surface was achieved by connecting anode plate and grid
together [see Fig.~\ref{roehre}].
\end{thebibliography}
\end{document}